# Berezinskii-Kosterlitz-Thouless transition in a photonic lattice

Guohai Situ[1,2], Stefan Muenzel[1], and Jason W. Fleischer[1]

[1]*Department of Electrical Engineering, Princeton University, Princeton, NJ 08540, USA*

[2] *Laboratory of Information Optics and Opto-Electronic Technology, Shanghai Institute of Optics and Fine Mechanics, Chinese Academy of Sciences, Shanghai, 201800, China*

**Abstract**

**Phase transitions give crucial insight into many-body systems, as crossovers between different regimes of order are determined by the underlying dynamics. These dynamics, in turn, are often constrained by dimensionality and geometry. For example, in one- and two-dimensional systems with continuous symmetry, thermal fluctuations prevent the formation of long-range order[1,2]. Two-dimensional systems are particularly significant, as vortices can form in the plane but cannot tilt out of it. At high temperatures, random motion of these vortices destroys large-scale coherence. At low temperatures, vortices with opposite spin can pair together, cancelling their circulation and allowing quasi-long-range order to appear. This Berezenskii-Kosterlitz-Thouless (BKT) transition[3,4] is essentially classical, arising for example in the traditional XY model for spins, but to date experimental evidence has been obtained only in cold quantum systems. Measurements of superfluid sound speed[5] and critical velocity[6] have been consistent with scaling predictions, and vortices have been observed directly in cold atom experiments[7,8]. However, the presence of trapping potentials restricts measurement to vortex density, rather than number, and obscures the process of vortex unbinding. Further, atom and fluid experiments suffer from parasitic heating and difficulties in phase recording, leading to results that differ from theory in many quantitative aspects. Here, we use a nonlinear optical system to directly observe the ideal BKT transition, including vortex pair dynamics and the correlation properties of the wavefunction, for both repulsive and attractive interactions (the photonic equivalent of ferromagnetic and antiferromagnetic conditions[9]). The results confirm the thermodynamics of the BKT transition and expose outstanding issues in the crossovers to superfluidity and Bose-Einstein condensation.**

It has been 40 years since Berezinskii[3], Kosterlitz and Thouless[4] discovered the phase transition that now bears their name. It is a remarkable transition in that it is topological: the dynamics of vortex pairs change the statistical properties of the fields without breaking the (continuous) symmetry of the system[2]. More specifically, unbinding of these pairs above a critical temperature changes the spatial correlation function from algebraic to exponential, with a crossover from superfluid behavior to that of a plasma-like gas of free (topological) charges. The BKT transition has much in common with Bose-Einstein condensation (BEC), which can also occur in 2D when the system is bounded[8,10]. On the other hand, BKT crossovers can appear in other contexts[3,4], leading to open questions about the phase space in which BEC and BKT dynamics exist[11]. Likewise, there are many issues within the BKT regime itself that remain unresolved.

To achieve a suitable competition between coherence and thermal fluctuations, all observations of BKT dynamics to date have utilized cold quantum systems. These span a variety of condensed matter systems, including superfluids[5], superconductors[13], exciton-polariton systems[12], and collisional[14] and cold[8] atoms, across a wide range of temperatures, confining potentials, and boundary conditions. Nevertheless, many fundamental features of the BKT transition remain obscure, including the number of vortices produced, a direct measurement of their correlation properties, and their behavior as a function of interaction sign and strength.

Optical systems provide a nearly ideal platform for studying these issues, as initial beam profiles can be controlled easily and their properties can be imaged directly. Coherent laser light provides a zero-temperature baseline, so that purely superfluid behavior can be observed without interference from a normal component[15,16], while thermal features can be controlled by adding a random-phase component to the beam. To date, thermalization and condensation of photons[17-19] has been observed, but no evidence for an optical BKT transition has been found.

Here, we observe the thermodynamics of vortices by considering the nonlinear propagation of a random-phase optical beam. As evolution of the beam in a homogeneous medium resulted in condensation only, we modify the conditions of Ref. 19 by introducing a weak lattice potential. This accomplishes three things. First, the periodicity breaks the phase-matching condition of four-wave mixing, suppressing the cascade mechanism of condensate formation[20]. Second, the band structure modifies the dispersion/diffraction relation[21], giving an effective mass to the photons and altering the stiffness of the corresponding fluid[22]. Third, the localized sites provide a direct mapping to the classical XY model from condensed matter physics[9,23], which is the paradigm description of BKT dynamics (see Methods for details).

The experimental setup is shown in Fig. 1. A finite-size, 10mW beam is projected onto a spatial light modulator (SLM). The SLM creates a random-phase field with a user-defined correlation length $l_c$. Nonlinear wave action occurs in a 5x5x10mm SBN:75 ($Sr_{0.75}Ba_{0.25}Nb_2O_6$) photorefractive crystal with a coupling strength controlled by applying a voltage across the c-axis[24]. A periodic potential is optically induced by interfering plane waves in the crystal[25], while a separate reference beam is split off from the input laser to measure relative phase. This latter beam gives the location and sign of vortices through the characteristic fork pattern of the singularity.

The photonic system inherits all the properties of the 2D XY model. In particular, there is a critical temperature $T_c$ at which vortex pairs unbind and separate into a gas of free charges. This occurs throughout the crystal as the light propagates, but only the output face of the crystal can be imaged. Nevertheless, a "snapshot" at the propagation distance z = $L_{crystal}$ (Fig. 1b) clearly shows the main ingredients: 1) many more vortices at the output compared with the input, 2) a set of vortex pairs, and 3) a set of free vortices. The number of free vortices, as well as their distribution, determines the superfluid properties of the light. These quantities are plotted in Figs. 2 and 3, respectively.

The statistical dynamics of the system can be characterized by the correlation function $C(r) = \iint_{r \leq \sqrt{x^2+y^2} \leq r+dr} dxdy \langle \psi^*(x_0, y_0)\psi(x_0 + x, y_0 + y) \rangle$, where $\psi$ is the optical wavefunction. Above the transition, vortex creation and unbinding destroys coherence, resulting in an exponentially decaying correlation function. The correlation length $\xi(T) \sim \exp[B(T - T_c)^{-1/2}]$ has the typical $(T - T_c)^{1/2}$ dependence expected from Ginzburg-Landau theory but with an essential singularity arising from (renormalized) vortex interactions [26]. On average, the correlation length determines the density of vortices in a given area $\Omega$, through $N/\Omega \sim \xi^{-2}$, so that the expected number of vortices is[9,26]

$$N(T) = AH(T - T_c) \exp[-2B(T - T_c)^{-1/2}] \quad (1)$$

where $H(T)$ denote the Heaviside step function and $A$ and $B$ are non-universal fitting parameters related to the integration area (system size). As shown in Figs. 2a,b, the number of new vortices is well-matched by the prediction (1) ($R^2 > 0.995$), with a constant value $A = 11000$ for all cases. The experiments confirm that the transition is the same for attractive or repulsive interactions, but the data was too noisy to observe the weak logarithm dependence of the transition temperature with interaction strength[27]. In all cases, the number of vortices produced is lower in the self-focusing case, due to enhanced attraction and recombination of vortex pairs.

The average separation distance between positive and negative vortices is shown in Fig. 3. For low initial kinetic energy, below the transition temperature, there are no new vortices produced, and the output PDF matches that of the input (with higher variance at the output). At the transition temperature, there is a clear difference in distributions, with the average separation distance increasing after evolution (propagation) through the crystal. For self-focusing nonlinearity, there is also peak showing decreased separation, due to increased vortex attraction. Well above transition, these peaks remain, but there is a change of slope in the long-range tails showing a reduced number of vortex pairs with large separation distance. This fall-off indicates an effective screening between elements of the vortex gas.

It proves convenient to express the spatial distribution in terms of the correlation parameter $\langle|A_\Omega|\rangle$, which is defined through $\langle|A_\Omega|^2\rangle \propto \Omega \int_0^{\sqrt{\Omega}} r|C(r)|^2 \, dr$ [9,28]. Above the transition temperature, the exponential decay gives $\langle|A_\Omega|^2\rangle \propto \Omega$. Below the transition, the field should have quasi-long-range order, with a scale-free correlation function $C(r) \propto r^{-\alpha(T)}$ that gives $\langle|A_\Omega|^2\rangle \propto \Omega^{2-\alpha}$ [26]. Hence, the crossover can be characterized by fitting $\langle|A_\Omega|\rangle$ to $\Omega^{\kappa(T)}$, where $\kappa = 0.5$ above the transition and $\kappa = 0.875$ at the transition temperature $T_c$ ($\alpha_c = 0.25$ as $T \to T_c$ from below)[26,29]. Experimental measurement of this exponent is shown in Figs. 2c,d. It is clear that the correlation functions obey the exact scaling behavior predicted by BKT theory, but the predicted step function at $T_c$ is replaced by a smooth transition, an effect attributable to the finite size of our system[8,30]. Saturation of the exponent at low temperatures (rather than a continuation to $\kappa=1$) was observed in Ref. 8 as well and remains an outstanding issue. We note that the exponent $\kappa$ gives the inverse superfluid stiffness, through $2\pi[2 - 2\kappa(T)]$; its change across transition corresponds to a universal jump in density[29]. This jump, which compounds the existing compressibility of the light[15], is difficult to observe directly.

Theoretically, the BKT transition retains the continuous symmetry of the system (it is a topological transition), and a lattice potential is not necessary for its dynamics. Experimentally, we find that the lattice is necessary, both for vortex production and as a stabilizing influence. Tunneling between sites gives an additional parameter when compared with the homogeneous case, while the band structure modifies the density of states and allows vortices to be stable under self-focusing nonlinearity[21]. The relation between the lattice and the critical temperature can be seen from the free energy expression F = E – TS = ($\pi J - 2T$)ln($L/a$), where $J$ is the (spin-spin) coupling between sites, and $L/a$ gives the ratio between the system size and the vortex core radius. When the lattice spacing is changed, the tunneling rate $J$ is changed, corresponding to a change in the transition temperature (Fig. 4a). To maintain a constant density jump at the critical temperature, there must be an inverse relationship between $T_c$ and $J$. From an entropy point of view, there are more sites available for occupation (for a fixed area) as the lattice spacing decreases, so the critical temperature must decrease to maintain $E_c \sim T_c S$. Interestingly, there is a finite transition temperature as the lattice spacing → 0. In this limit, thermal excitations see a potential which is effectively uniform, reducing the dynamics to that of a homogeneous system. Experiments in this regime reveal only wave condensation, with no evidence of vortex production[19].

It is difficult to compare condensation and BKT dynamics directly in the same optical experiment, as the finite length of the crystal (limited evolution time) necessitates initial conditions that are most favorable for each transition. For the former, it is a Gaussian distribution in a homogeneous system, while for the latter it is an exponential distribution in a lattice. Nevertheless, the crystal response is the same, and a comparison can be made when the parameters of the input beam are normalized to $E/N = \int k^2 n(k) dk / \int n(k) dk$, where $n(k)$ is the initial power spectrum $n(\mathbf{k})\delta(\mathbf{k} - \mathbf{k'}) = \langle \widetilde{\psi}^*(\mathbf{k}, z=0) \widetilde{\psi}(\mathbf{k'}, z=0) \rangle$. As shown in Fig. 4b, the respective figures of merit (condensate fraction and number of unbound vortices) have a common crossover energy. That is, each experiment shows the appearance of system-wide order at the same temperature, supporting predictions that Bose-Einstein condensation is the low-energy limit of BKT dynamics[11]. On the high-temperature side, the disappearance of long-range order results from two different mechanisms: smooth phase variations (phonons) in the BEC experiment and singular perturbations (vortices) in the BKT case. Remarkably, both types of excitation give the same scaling behaviour: the Rayleigh-Jeans distribution that describes the final condensed state and its thermal cloud (the classical limit of the BE distribution) is a Lorentzian power spectrum, which also gives an exponentially decreasing correlation function. Indeed, it is well known that long-range scaling is a coarse measure of disorder, and that vortices are the distinguishing feature of the BKT transition[4].

The crossover between BEC and BKT dynamics requires a more detailed study of critical point stability and the nature of thermal excitations. For interacting waves, wave turbulence theory has successfully described the non-equilibrium approach to thermalization and condensation[19]. To date, however, the theory has had trouble accommodating vortices, both because they are singular structures and because they are not weak perturbations. Nevertheless, the coupling of vortices with pressure variations is fundamental to (quantum) instability[16] and turbulence. The results here give a thermodynamic endpoint to these studies, as well as a starting point for exploring the full phase space of non-equilibrium and turbulent dynamics.

This work was supported by the Air Force Office of Scientific Research.

## Methods

In the tight-binding approximation, the diffraction operator becomes discretized, so that the parxial evolution of the wavefunction $\psi$ may be written[23]

$$i\frac{d\psi_n}{dz} = -J\sum_{m(n)} \psi_m + \gamma|\psi_n|^2\psi_n$$

where $\psi_n = |\psi_n|\exp[i\theta_n]$ is the complex wavefunction of the mode in the nth lattice site, the phase $\theta_n \in [0,2\pi)$ corresponds to a localized "spin," $J$ is the coupling coefficient between nearest neighbors, the sum $\Sigma_{m(n)}$ indicates a sum over all nearest neighbors of the nth lattice site, and $\gamma$ is the nonlinear coefficient. When the nonlinearity is strong enough the Hamiltonian $\widehat{H}(\psi_n) = -\frac{1}{2}\sum_{<m,n>}(\psi_n\psi_m^* + \psi_n^*\psi_m) + \frac{1}{2}|\Gamma|\sum_{n=1}^{N} I_n^2$, can be decoupled into two parts[9]

$$H_{amplitude} = \frac{1}{2}|\Gamma|\sum_{n=1}^{N} I_n^2$$

$$H_{phase} = -2\sum_{\langle m,n\rangle} \cos(\theta_n - \theta_m)$$

where $\Gamma = \gamma|\psi_n|^2/J = \gamma I_n/J$ is the ratio of self-energy to coupling strength. The angular part $H_{phase}$ is identical to the Hamiltonian of the traditional XY model of condensed matter physics. The amplitude part $H_{amplitude}$ does not contain coupling terms between adjacent lattice sites and therefore can be solved for the intensity distribution function, which reads $P(I) = \sqrt{\frac{|\Gamma|}{2\pi T}}\exp\left[-\frac{|\Gamma|}{2T}(I-1)^2\right]$. By analogy with thermodynamics, $T$ is the effective temperature of the system.

Experimentally, the signal beam is a 10mW laser beam at 532nm, extraordinarily polarized, with an input profile $P$ made using a spatial light modulator. Different temperatures are obtained by spatially filtering the random phases with a low-pass filter of controllable width, which is related to the correlation length. Ensemble-averages are obtained using 5-10 different representations of $P$ for each data point. A photonic lattice is optically induced by interfering plane waves that are ordinarily polarized[25]. Nonlinear interactions are controlled by applying a an electric field $E_0$ across a 5x5x10mm SBN:75 crystal, whose nonlinear index change $\Delta n = -\frac{1}{2}n_0^3 r_{ij} E_0 I$, where $n_0 = 2.3$, $r_{ij}$ is the electro-optic coefficient and $I = |\psi|^2$ is the intensity formed from interfering modes[24]. At the maximum applied voltage of $\pm 450$V, the nonlinear index change is measured to be $|\Delta n/n_0| = 6\cdot 10^{-4}$. At the output, phase is measured by interfering the signal beam with a reference plane wave.

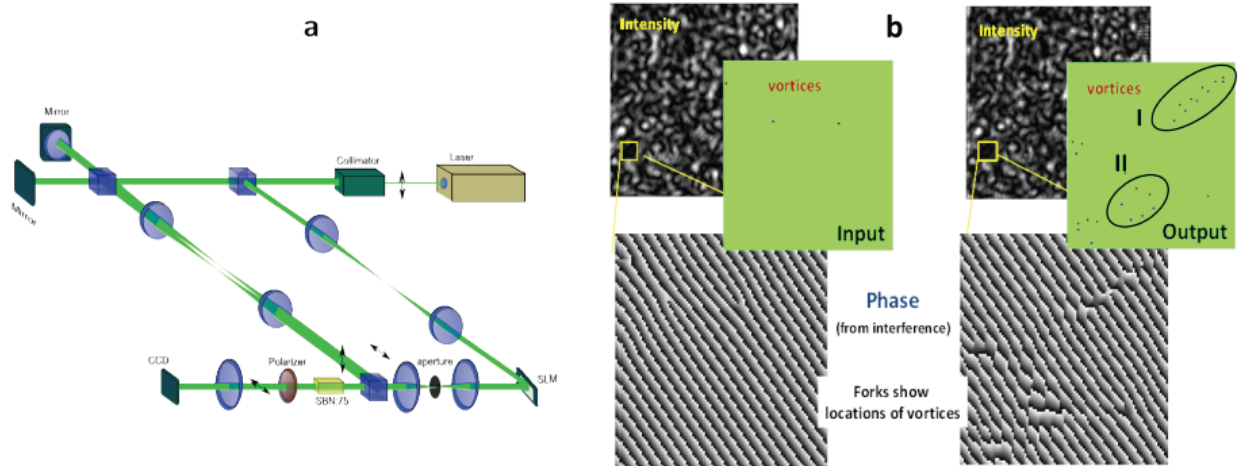

**Figure 1. Experimental setup and typical results**. (a) Light from a 532 nm laser is split into two beams by using a beam splitter. One beam creates an interference grating on an 5×5×10 mm SBN:75 photorefractive crystal. The other beam acquires a user-controlled random-phase pattern from a spatial light modulator, which is then imaged onto the input plane of the crystal. The output plane of the crystal is imaged onto the CCD plane and interferes with a tilted reference beam (not shown). (b) Typical input and output pictures, showing a proliferation of vortices (forks in the phase) above the critical temperature. The labels highlight sets of bound (I) and free (II) vortices.

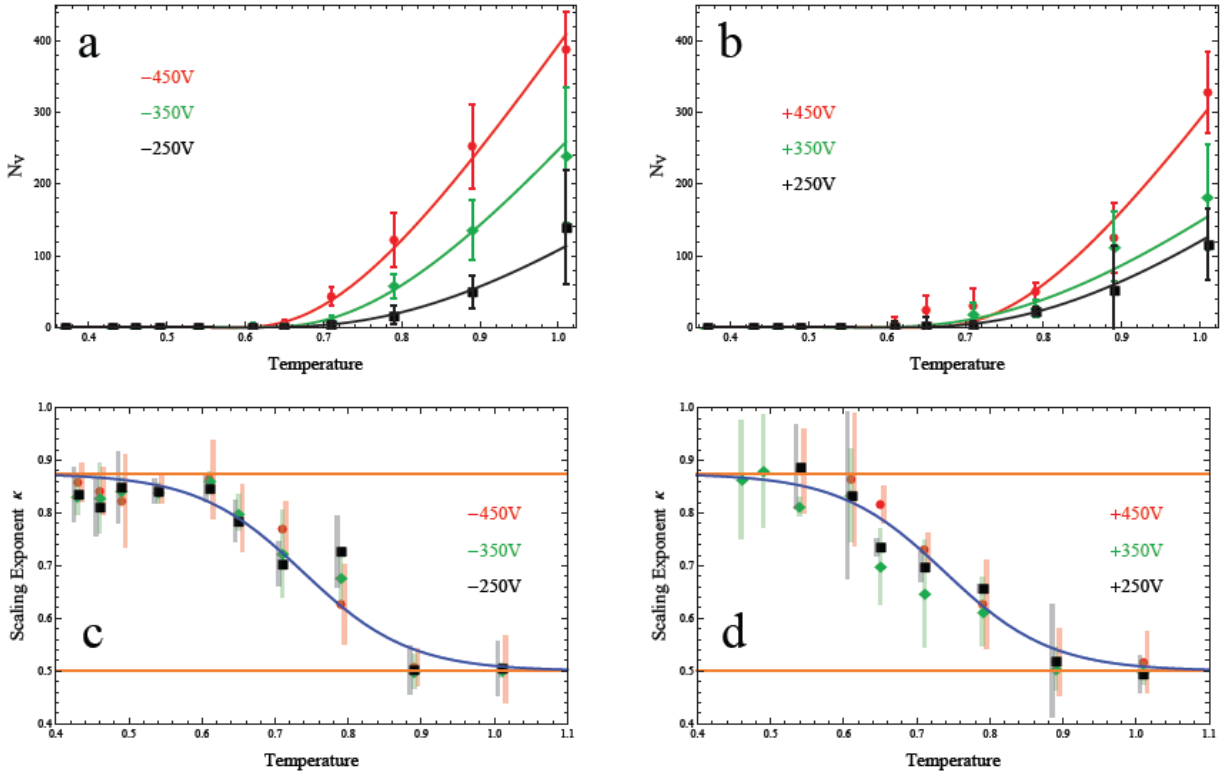

**Figure 2. Direct measurement of vortex number and correlation function**. (a,b) Number of unbound vortices for (a) self-defocusing (repulsive) and (b) self-focusing (attractive) nonlinearity, as a function of interaction strength and input temperature. Solid lines indicate predictions from BKT theory. (c,d) Scaling exponent κ vs. temperature. When κ = 0.5, the correlation function decays exponentially, while when κ = 0.875 it decays algebraically. The exponent κ gives the inverse superfluid stiffness through $2\pi[2 - 2\kappa(T)]$. The solid lines are a fit to a *tanh* function.

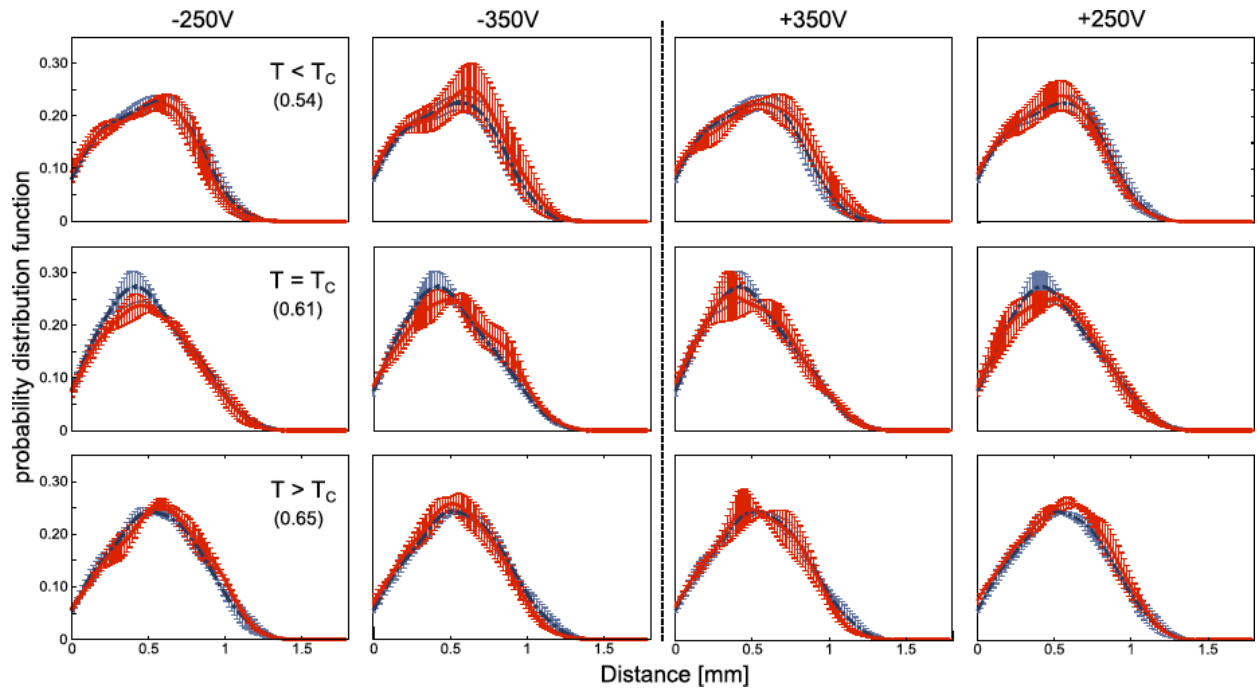

**Figure 3. Distribution of separation distances for vortex pairs.** Below transition, the output probability distribution function (PDF) is the same as the input PDF, with higher variance. At transition, a new peak appears at larger separation distances, signifying the unbinding of vortex pairs. There is also a shift in the main peak towards smaller distances (esp. in the self-focusing case), indicating enhanced vortex attraction. Above transition, there is a distinct change in the PDF for large separation distances.

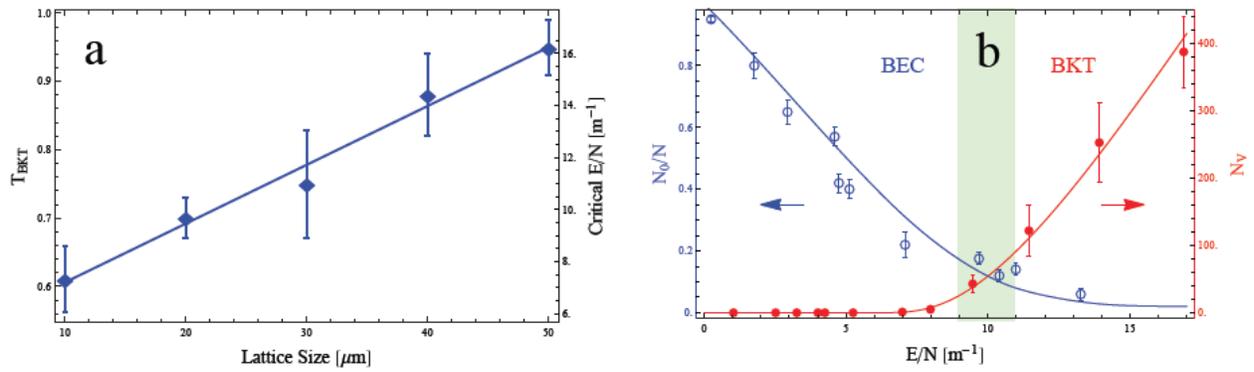

**Figure 4. BEC-BKT crossover.** (a) The critical temperature for vortex production as a function of lattice period. (b) Comparison of photon condensation in a homogenous system [19] and BKT transition in a lattice (Fig. 2a), as a function of the initial kinetic energy. Though the setups are different, there is a common crossover to system-wide order.